\documentclass{emulateapj}
\usepackage{graphics}
\usepackage{psfig}
\usepackage{ulem}
\usepackage{longtable}

\def\sn{{SN~2006jc}}

\def\C{{\sl Chandra}}

\def\S{{\sl Swift}}

\def\ls{\mathrel{\mathchoice {\vcenter{\offinterlineskip\halign{\hfil
$\displaystyle##$\hfil\cr<\cr\sim\cr}}}
{\vcenter{\offinterlineskip\halign{\hfil$\textstyle##$\hfil\cr
<\cr\sim\cr}}}
{\vcenter{\offinterlineskip\halign{\hfil$\scriptstyle##$\hfil\cr
<\cr\sim\cr}}}
{\vcenter{\offinterlineskip\halign{\hfil$\scriptscriptstyle##$\hfil\cr
<\cr\sim\cr}}}}}

\begin{document}

\title{{\sl Swift} and {\sl Chandra} Detections of Supernova 2006\lowercase{jc}: \\
Evidence for Interaction of the Supernova Shock with a Circumstellar Shell}

\author{S.~Immler\altaffilmark{1,2,3},
M.\ Modjaz\altaffilmark{4},
W.\ Landsman\altaffilmark{1},
F.\ Bufano\altaffilmark{1,5},
P.\ J. Brown\altaffilmark{6}, 
P.\ Milne\altaffilmark{7}, 
L.\ Dessart\altaffilmark{7},
S.\ T.\ Holland\altaffilmark{1,2,8},
M.\ Koss\altaffilmark{1,3},
D.\ Pooley\altaffilmark{9,10},
R.\ P.\ Kirshner\altaffilmark{4},
A.\ V.\ Filippenko\altaffilmark{9},
N.\ Panagia\altaffilmark{11},
R.\ A.\ Chevalier\altaffilmark{12},
P.\ A.\ Mazzali\altaffilmark{13,14},
N.\ Gehrels\altaffilmark{1},
R.\ Petre\altaffilmark{1}, 
D.\ N.\ Burrows\altaffilmark{6}, 
J.\ A.\ Nousek\altaffilmark{6}, 
P.\ W.\ A.\ Roming\altaffilmark{6},
E.\ Pian\altaffilmark{13},
A.\ M.\ Soderberg\altaffilmark{15}, and
J.\ Greiner\altaffilmark{16}
}

\email{stefan.immler@nasa.gov}
\altaffiltext{1}{Astrophysics Science Division, NASA Goddard Space Flight Center,
Greenbelt, MD 20771.}
\altaffiltext{2}{Center for Research and Exploration in Space Science and 
Technology, NASA/GSFC, Greenbelt, MD 20771.}
\altaffiltext{3}{Department of Astronomy, University of Maryland, College
Park, MD 20742.} 
\altaffiltext{4}{Harvard-Smithsonian Center for Astrophysics, 60 Garden Street, 
Cambridge, MA 02138.}
\altaffiltext{5}{INAF-Osservatorio Astronomico, vicolo dell'Osservatorio 5, 
I-35122 Padova, Italy.}
\altaffiltext{6}{Department of Astronomy and Astrophysics, Pennsylvania 
State University, 525 Davey Laboratory, University Park, PA 16802.}
\altaffiltext{7}{Steward Observatory, 933 North Cherry Avenue, RM N204 Tucson, 
AZ 85721.}
\altaffiltext{8}{Universities Space Research Association,
10211 Wincopin Circle, Columbia, MD 21044.}
\altaffiltext{9}{Department of Astronomy, University of California, 
Berkeley, CA 94720-3411.}
\altaffiltext{10}{Chandra Fellow.}
\altaffiltext{11}{Space Telescope Science Institute, 3700 San Martin Dr., 
Baltimore, MD 21218.}
\altaffiltext{12}{Department of Astronomy, University of Virginia, 
P.O. Box 400325, Charlottesville, VA 22904.}
\altaffiltext{13}{INAF, Osservatorio Astronomico di Trieste, via G.B. 
Tiepolo 11, 34131 Trieste, Italy.}
\altaffiltext{14}{Max-Planck-Institut f\"ur Astrophysik, Karl-Schwarzschild
Strasse 1, 85741 Garching, Germany.}
\altaffiltext{15}{Department of Astronomy, 
California Institute of Technology, MS 105-24, Pasadena, CA 91125.}
\altaffiltext{16}{Max-Planck-Institut f\"ur Extraterrestrische Physik, 
Giessenbachstrasse, 85748 Garching, Germany.}

\shorttitle{{\sl Swift} and {\sl Chandra} Detections of Supernova 2006\lowercase{jc}}
\shortauthors{Immler et~al.}

\begin{abstract}
The peculiar Type Ib supernova (SN) 2006jc has been observed with the
UV/Optical Telescope (UVOT) and X-Ray Telescope (XRT) on board the \S\ observatory 
over a period of 19 to 183 days after the explosion. Signatures of interaction
of the outgoing SN shock with dense circumstellar material (CSM) are detected, 
such as strong X-ray emission ($L_{0.2-10} > 10^{39}~{\rm erg~s}^{-1}$) and the 
presence of Mg~II 2800~\AA\ line emission visible in the UV spectra.
In combination with a \C\ observation obtained on day 40 after the explosion,
the X-ray light curve is constructed, which shows a unique rise of the X-ray emission
by a factor of $\sim 5$ over a period of $\sim 4$ months, followed by a rapid decline.
We interpret the unique X-ray and UV properties as a result of the SN shock 
interacting with a shell of material that was deposited by an outburst of the 
SN progenitor two years prior to the explosion. 
Our results are consistent with the explosion of a Wolf-Rayet star that 
underwent an episodic mass ejection qualitatively similar to those of
luminous blue variable stars prior to its explosion. This led to the formation of a 
dense ($\sim 10^{7}~{\rm cm}^{-3}$) shell at a distance of $\sim 10^{16}$~cm from the 
site of the explosion, which expands with the WR wind at a velocity of 
$(1300 \pm 300)~{\rm km~s}^{-1}$.
\end{abstract}


\begin{figure*}[t!]
\unitlength1.0cm
    \begin{picture}(8.7,6) 
\put(0,0){ \begin{picture}(8.7,6)
	\psfig{figure=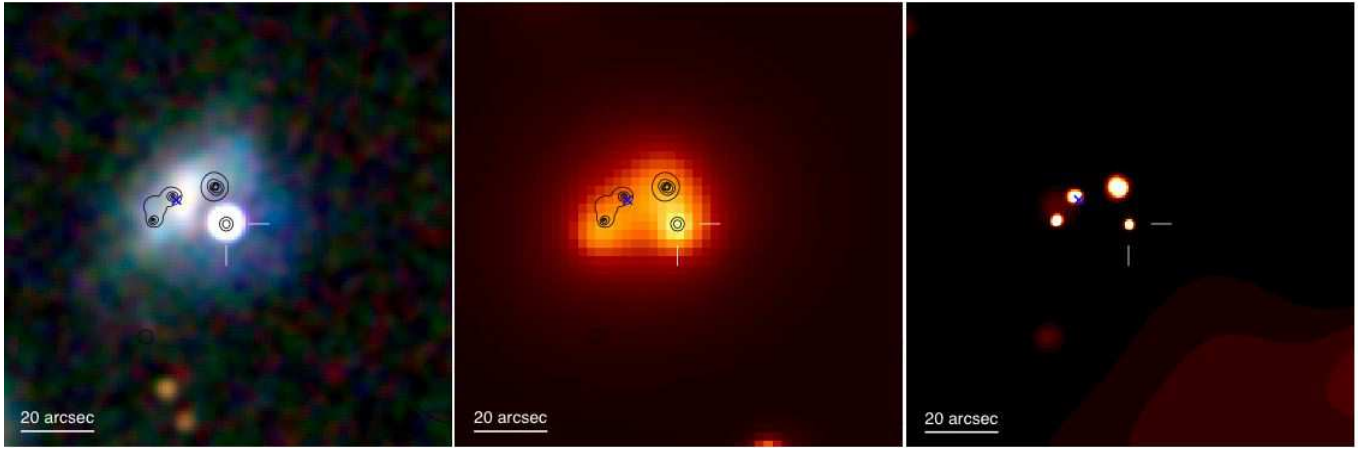,width=18cm,clip=}
	\end{picture}
	}
    \end{picture}
\caption{
{\sl Swift} UVOT optical (left), XRT X-ray (middle), and \C\ X-ray (right) images of 
\sn\ and its host galaxy UGC~4904. 
{\bf Left-hand panel:}
The \S\ optical image was constructed from the UVOT $V$ (288~s exposure time; red), 
$B$ (357~s; green), and $U$ (432~s; blue) filters obtained on 2006 October 16.
The optical position of \sn\ (Nakano et~al.\ 2006) is 
indicated by white tick marks; the nucleus of the host galaxy is marked by a blue cross.
The size of the image is $2' \times 2'$. 
{\bf Middle panel:} 
X-ray image (0.2--10~keV band) from the merged 150~ks \S\ XRT data obtained between 
2006 October 13 and 2007 March 3.
The image is adaptively smoothed to archive a signal-to-noise ratio (S/N) in the range 2.5--4.
{\bf Right-hand panel:} The 0.2--10~keV X-ray image was constructed from 
the 10.0~ks \C\ ACIS data obtained on 2006 November 4 and is adaptively smoothed to 
archive a S/N in the range 2.5--4. The contour levels 
($0.01$, $0.1$, $1$, $10$, and $100 \times 10^{-4}$~counts~pixel$^{-1}$) 
are overlaid in black on the \S\ optical and X-ray images.
}
\label{fig1}
\end{figure*}

\section{Introduction}
\label{introduction}

\sn\ was discovered on 2006 October 9.75 (all times are UT)
with an apparent magnitude of 13.8 in unfiltered CCD exposures (Nakano et~al.\ 2006).
No object was visible at the position of the SN on 2006 September 22 (limiting 
magnitude 19.0). Subsequent observations showed that the SN was around peak at the 
time of discovery (October 10.33: unfiltered magnitude 13.8, Nakano et~al.\ 2006; 
October 13.67: \S\ $V$-band magnitude 14.3, Brown, Immler \& Modjaz 2006).
Throughout this {\it Letter} we adopt an estimated explosion date around 
September 25 ($\pm5$ days). 
Approximately two years earlier (2005 October) a variable object ($\sim 18$~mag) 
was discovered close to the position of \sn\ ($\ls 0\farcs3$ offset; Pastorello et~al.\ 2007)
which is thought to be associated with \sn.

Type Ib/c SNe (see Filippenko 1997 for a review) are the result of the 
core collapse of a hydrogen-deficient, massive star (e.g., a Wolf-Rayet 
star); its outer layers were stripped by either mass transfer to a 
companion or by a strong stellar wind. Spectra of SNe Ib suggest 
that the progenitors have lost most of the H envelope, while 
progenitors of SNe~Ic have lost the H layer and much of the He layer. 
The gas lost from the progenitors can be shock heated by the outgoing
SN shock, giving rise to non-thermal radio, UV, and thermal X-ray emission.

In this {\sl Letter} we present and discuss the optical, UV (with emphasis on the 
UV grism data), and X-ray emission properties of the peculiar Type Ib \sn\ as observed
by \S, and X-ray emission as observed by \C.
A distance of 24~Mpc is adopted throughout the paper ($z=0.00557$, Nordgren et~al.\ 1997).
A more detailed discussion of the near-IR/optical/UV light curve of \sn\ will be
presented in a separate study.

\section{Observations}
\label{obs}

\subsection{{\sl Swift} UVOT Optical/UV Observations}
\label{uvot}
The Ultraviolet/Optical Telescope (UVOT; Roming et~al.\ 2005) and X-Ray Telescope
(XRT; Burrows et~al.\ 2005) on board the \S\ Observatory (Gehrels et~al.\ 2004) 
began observing \sn\ on 2006 October 13.67. 
The HEASOFT\footnote{http://heasarc.gsfc.nasa.gov/docs/software/lheasoft/} 
(v6.3) and \S\ Software (v2.7, build 21) tools and latest calibration products 
were used to analyze the data.

The SN was detected with UVOT at $\alpha = 9^{\rm h}17^{\rm m}20^{\rm s}.81$, 
$\delta = +41^\circ54'32\farcs9$ (equinox 2000.0), consistent with the position
reported by Nakano et~al.\ (2006), with magnitudes of
$V = 14.3$, 
$UVW1$ [216--286~nm FWHM] $= 13.2$, 
$UVM2$ [192--242~nm] $= 13.5$, and 
$UVW2$ [170--226~nm] $= 13.7$.
Statistical and systematic errors are 0.1~mag each.

Forty-eight individual exposures were obtained between 2006 October 13.67, 
and 2007 March 3.28.
Assuming an explosion date of 2006 September 25, the observations correspond to days 
19--183 after the explosion.
A \S\ $V$-, $B$-, and $U$-filter image of \sn\ and its host galaxy UGC~4904 
(Hubble type SB; NED) is shown in the left-hand panel of Fig.~1. 

\subsection{{\sl Swift} Grism Observations}
\label{grism}
The UVOT includes two grisms which operate in the visible and UV bands. 
The UV grism has a nominal wavelength range of 1800--2900~\AA, while the $V$ 
grism has a wavelength range of 2800--5200~\AA.
We obtained three UV grism and three $V$ grism spectra of \sn. 
Due to contamination from the host galaxy and bright stars in the field we only 
use three uncontaminated UV (days 21, 25, and 28 after the explosion) and one 
$V$ grism spectrum (day 40) in this study.
The wavelength has an uncertainly of up to 20~\AA, caused by uncertainties in the 
dispersed zeroth-order spectrum scale. The UVOT UV grism range is nominally 
1800--2900~\AA, but extends up to 4900~\AA\ with a lower sensitivity
than the $V$ grism and the possibility of contamination by the second-order 
overlap. We therefore selected a wavelength range from 2000~\AA\ to 4000~\AA.
The extracted spectra, shown in Fig.~2, were smoothed using a running 
average of 3 points ($\sim 8$~\AA).
The synthetic magnitudes, measured by folding the spectra through the UVOT $UVW1$ 
filter transmission curve, match the photometric magnitudes within the uncertainties. 

\subsection{{\sl Swift} XRT Observations}
\label{xrt}

The \S\ XRT observations were obtained simultaneously with the UVOT observations.
X-ray counts were extracted from a circular region with a 3~pixel ($7\farcs2$) 
radius centered on the optical position of the SN. The background was extracted 
locally from a source-free region of radius of $30''$ to account for detector and 
sky background, and for diffuse emission from the host galaxy.
The superior resolution \C\ image (see below) was smoothed to the larger 
point spread function of the \S\ XRT ($15''$ half power diameter) and the XRT count 
rates were corrected for residual contamination with a nearby ($10''$) and variable 
X-ray source. A \S\ XRT X-ray image of \sn\ and its host galaxy is shown in the 
middle panel of Fig.~1.

\subsection{{\sl Chandra} Observation}
\label{xmm}
A \C\ Advanced CCD Imaging Spectrometer (ACIS) Director's Discretionary Time
(DDT) observation was performed on 2006 November 4.33 (PI Immler, sequence 500815), 
corresponding to day 40 after the explosion. 
CIAO\footnote{http://cxc.harvard.edu/ciao/} (v3.4),
FTOOLS\footnote{http://heasarc.gsfc.nasa.gov/docs/software.html} (v6.3)
and the latest \C\ calibration products were used to analyze the data.
Inspection of the ACIS data for periods with a high particle background showed 
no contamination of the data, which resulted in a clean exposure time of 10.0~ks.
The \C\ ACIS image of the \sn\ field is given in the right-hand panel of Fig.~1.

\section{Results and Discussion}
\label{results}

\begin{figure}[t!]
\unitlength1.0cm
    \begin{picture}(6,6)
\put(-0.7,-0.7){ \begin{picture}(6,6)
	\psfig{figure=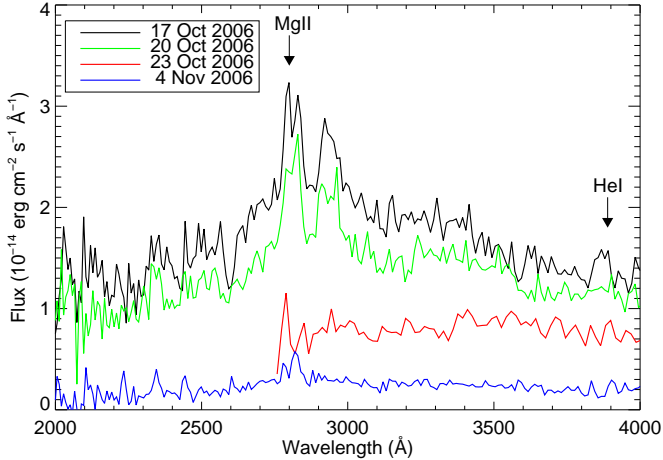,width=9.5cm,clip=}
	\end{picture}
	}
    \end{picture}
\vspace{6mm}
\caption{
{\sl Swift} grism observations of SN 2006jc obtained
on 2006 October 17 (UV grism), 20 (UV), 23 ($V$), and November 4 (UV), 
corresponding to days 21, 25, 28, and 40 after the explosion.
}
\label{fig2}
\end{figure}

\subsection{Optical/UV Emission}
\label{uvotresults}

\sn\ appears to be a peculiar object in the optical range with features similar 
to those of both Type IIn and Type Ib SNe (Foley et~al.\ 2007; Pastorello et al.\ 2007;
Smith et~al.\ 2007). 
The intermediate-width He lines of \sn\ are similar to the narrow H lines seen in 
Type IIn SN spectra (Schlegel 1990; Chugai et~al.\ 1994; Filippenko 1997) and can be 
understood as the result of an interaction between the outgoing SN shock and a dense, 
H+He-rich CSM.
A strong Mg~II emission line at 2800~\AA\ and possibly Fe~II (between 2900--3000~\AA)
and a He~I 3889~\AA\ emission line (seen more clearly in ground-based spectra) 
are visible in the \S\ grism data (see Fig.~2).
Each of the emission lines become weaker with time.
The Mg~II emission line might be strong as Mg~II is one of the main coolants in 
the \sn\ shell. It is further overlapped by Mg~II absorption from both
Galactic and host-galaxy interstellar gas (Panagia et~al.\ 1980).
The most likely origin for the Mg~II and He~I emission lines is a UV-emitting shell 
consisting of gas originally ejected by the progenitor, likely associated with the 
luminous outburst found two years earlier. 

The later \S\ grism spectra ($>25$~d) are of lower quality but show how the
continuum becomes fainter with time. The UV flux deficiencies of SNe stem from the
cooling of the emitting region and from metal line blanketing (e.g., Hillier \& Miller 1998). 

\begin{figure}[t!]
\unitlength1.0cm
    \begin{picture}(6.8,6.8) 
\put(-1,0){ \begin{picture}(6.8,6.8)
	\psfig{figure=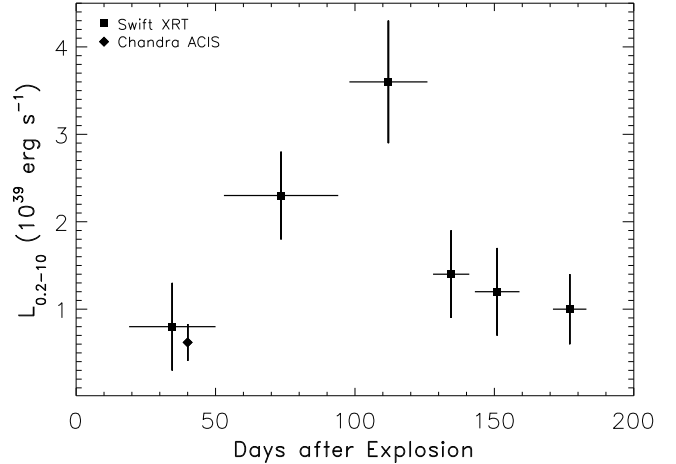,width=9.8cm,angle=0,clip=}
	\end{picture}
	}
    \end{picture}
\caption{
X-ray light curve (0.2--10~keV) of SN~2006jc as observed with the \S\ 
XRT and \C\ ACIS-S instruments. The time is given in days after the outburst.
Vertical error bars are statistical $1\sigma$ uncertainties; 
horizontal error bars indicate the periods covered by the observations 
(which are not contiguous).
} 
\label{fig3}
\end{figure}

\begin{deluxetable}{ccccccc}
\tabletypesize{\footnotesize}
\tablecaption{X-Ray Observations of SN~2006\lowercase{jc} \label{tab3}}
\tablewidth{0pt}
\tablehead{
\colhead{$t-t_0$} &
\colhead{Inst} &
\colhead{Exp} &
\colhead{Rate} &
\colhead{$f_{\rm x}$} &
\colhead{$L_{\rm x}$} &
\colhead{$\langle {\rm E} \rangle$} \\
\colhead{} &
\colhead{} &
\colhead{[ks]} &
\colhead{[$10^{-3}$]} &
\colhead{[$10^{-14}$]} &
\colhead{[$10^{39}]$} &
\colhead{[keV]} \\
\noalign{\smallskip}
\colhead{(1)}  &
\colhead{(2)}  &
\colhead{(3)}  &
\colhead{(4)}  &
\colhead{(5)}  &
\colhead{(6)}  &
\colhead{(7)}
}
\startdata
40	 & ACIS & 10.0 & $1.0\pm0.3$ & $0.8\pm0.3$ & $0.6\pm0.2$  & 5.0 \\
19--50	 & XRT  & 34.6 & $0.3\pm0.2$ & $1.2\pm0.8$ & $0.8\pm0.5$ & 3.9 \\
53--94	 & XRT  & 29.5 & $0.9\pm0.2$ & $3.3\pm0.7$ & $2.3\pm0.5$ & 3.4 \\
98--126	 & XRT & 31.2 & $1.5\pm0.3$ & $5.2\pm1.0$ & $3.6\pm0.7$ & 2.8 \\
128--141 & XRT  & 29.3 & $0.6\pm0.2$ & $2.0\pm0.7$ & $1.4\pm0.5$ & 2.5 \\
143--159 & XRT  & 29.1 & $0.5\pm0.2$ & $1.7\pm0.7$ & $1.2\pm0.5$ & 2.4 \\
171--183 & XRT  & 35.4 & $0.5\pm0.2$ & $1.5\pm0.6$ & $1.0\pm0.4$ & 2.3
\enddata
\tablecomments{
(1)~Days after the explosion of \sn\ (2006 September 25);
(2)~Instrument used: \C\ ACIS or \S\ XRT;
(3)~Exposure time in units of ks;
(4)~0.2--10~keV net count rate in units of cts~ks$^{-1}$.
Errors are photon statistical $1\sigma$ uncertainties;
(5)~Unabsorbed (0.2--10~keV) X-ray flux in units of $10^{-14}~{\rm erg~cm}^{-2}~{\rm s}^{-1}$.
Fluxes were converted from count rates using the HEASARC tool PIMMS (v3.9e), a thermal plasma spectrum
with a temperature from column~7, and corrected for the Galactic foreground column density of 
$N_{\rm H} = 1.45 \times 10^{20}~{\rm cm}^{-2}$ (Dickey \& Lockman 1990);
(6)~0.2--10~keV X-ray band luminosity in units of $10^{39}~{\rm erg~s}^{-1}$;
(7)~Mean energy of the X-ray photons. 
The mean energies of the photons in the \C\ ACIS and \S\ XRT cannot be compared 
directly because of differences in instrument responses and possible contamination of
the XRT data by nearby point sources.
}
\end{deluxetable}

\subsection{X-ray Emission}

\sn\ is detected in X-rays both with \S\ XRT and \C\ ACIS, at a luminosity of 
$\sim 10^{39}~{\rm erg~s}^{-1}$ at an early epoch ($<50$~d).
In order to study the temporal evolution of the X-ray emission with sufficient
photon statistics, we binned the XRT data into six consecutive time
intervals with exposure times around 30~ks each (see Table~1 and Fig.~3).
An increase of the X-ray flux by a factor of $\sim 5$ within the first $\sim 4$
months after the explosion is observed. Such an increase in the X-ray flux
is highly unusual for an X-ray emitting SN\footnote{See
http://lheawww.gsfc.nasa.gov/users/immler/supernovae\_list.html
for a complete list of X-ray SNe and references.}.
Although the limited photon statistics does not allow a detailed spectral 
characterization of the emission, we inspected the energy distribution of 
all X-ray photons that were recorded within a radius of 3 XRT image pixels ($7\farcs2$) 
centered on the optical position of the SN. The mean energy per time bin (listed in
column~7 in Table~1) shows a clear softening of the X-ray emission. 

\begin{figure}[t!]
\unitlength1.0cm
    \begin{picture}(6.9,6.9) 
\put(-0.4,0){ \begin{picture}(6.9,6.9)
	\psfig{figure=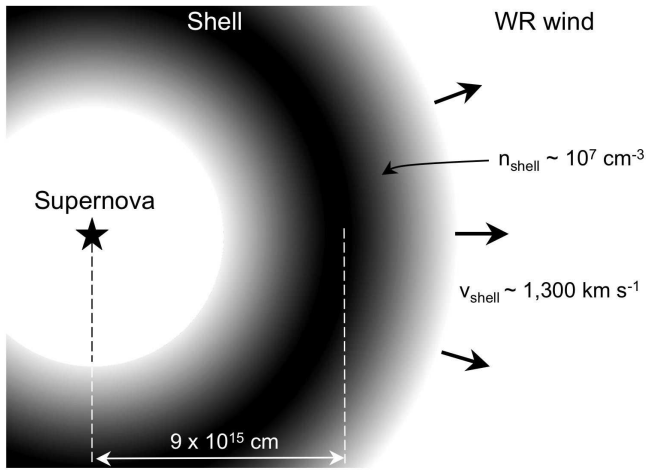,width=9.4cm,angle=0,clip=}
	\end{picture}
	}
    \end{picture}
\caption{
Illustration of the environment of SN~2006jc as inferred from the \S\ and
\C\ X-ray observations. The ejected shell is expanding at a speed of
$v_{\rm shell} = (1,300\pm300)~{\rm km~s}^{-1}$ with the wind blown-off by the
WR progenitor. At the time of the X-ray peak, around day 112 after the explosion 
of SN~2006jc, when the SN shock has traveled through the shell, the shell has 
reached a distance of $9\times10^{15}~{\rm cm}$ from the site of the explosion.
The density within the shell is $\sim 10^{7}~{\rm cm}^{-3}$.  
}
\label{fig4}
\end{figure}

A likely explanation for the brightening and subsequent fading in X-rays can 
be found in the history of the \sn\ progenitor: two years before the SN event,
a bright outburst ($M = -14.0$~mag) was noted at the position of the SN
(Nakano et~al.\ 2006). Foley et al.\ (2007) proposed that this outburst could 
have been the ejection of the outer layers of the progenitor in an outburst 
qualitatively similar to those experienced by luminous blue variables (LBVs; 
see Smith \& Owocki 2006).
The ejected shell would then be shock heated by the outgoing SN shock,
giving rise to the observed increase in X-rays as the SN
shock passes through the shell, followed by a decline of the X-ray emission
as the shock leaves the shell. This scenario is supported by a softening of the
X-ray emission over the observed period (see Table~1) as a result of the decreasing
absorbing column density as the SN shock passes through the dense shell.

Our X-ray measurements further allow a characterization of the physical properties
and geometry of the shell surrounding \sn. If the X-rays are produced by the 
shock-heated shell, the X-ray peak at $t \approx 112$~d after the explosion corresponds 
to a distance of the shell from the site of the explosion of 
$r = v_{\rm shock} \times t = 9 \times 10^{15}$~cm for a shock velocity of 
$v_{\rm shock} = 9,000~{\rm km~s}^{-1}$ (consistent with the FWHM of broad emission lines 
visible in the optical spectra; Pastorello et~al.\ 2007).
The X-ray luminosity of a thermal plasma is the product of the emission measure
and the plasma density within the emitting volume, $L_{\rm x} = \Lambda(T) {\rm d}V n^2$.
For a shell with a thickness of $\Delta R \sim 2 \times 10^{15}$~cm (estimated from 
the FWHM of the X-ray rise and decline of $\Delta t \sim 60$~days and 
$\Delta R = \Delta t \times v_{\rm shock}$), this can be rewritten as
$L_{\rm x} = 4\pi R^3 \frac{\Delta R}{R} \Lambda(T) n^2$. 
An effective (0.2--10~keV band) cooling function of 
$\Lambda = 3 \times 10^{-23}~{\rm erg~cm}^3~{\rm s}^{-1}$ for an optically thin thermal 
plasma with a temperature in the range $10^6$--$10^8$~K (Raymond, Cox \& Smith 1976) 
is adopted, which is consistent with the measured energies of the X-ray photons 
(see column~7 in Tab.~1).
Solving for the CSM number density of the shocked plasma, $n = \rho_{\rm csm}/m$, where 
$m$ is the mean mass per particle ($2.2\times10^{-24}$~g for a  fully ionized pure He plasma)
gives $n_{\rm csm} \sim 10^7~{\rm cm}^{-3}$ inside the X-ray emitting shell.
This leads to an estimate of the X-ray emitting mass that was lost by the progenitor
system over the two years period prior to the explosion of $\sim 0.01~{\rm M}_{\odot}$
(which represents a lower limit to the entire mass of the shell).

\section{Summary}
\label{summary}

The \S\ and \C\ observations of the peculiar Type Ib \sn\ reveal that the 
progenitor had a remarkable evolution and environment. The presence of Mg~II line 
emission and the rise of the X-ray emission over a period of $\sim 4$ months
after the explosion, followed by a fast decline, are likely due to the presence 
of a dense shell of material around the site of the explosion.
Our results clearly support the scenario proposed by Foley et~al.\ (2007) and Smith
et~al.\ (2007) that \sn\ was the explosion of a Wolf-Rayet star in a dense environment
which had a giant outburst before its core collapse (see Pastorello et~al.\ 2007 for a 
somewhat different interpretation involving a binary system).
The \S\ and \C\ observations allow, for the first time, a 
characterization of the physical properties and geometry of such a LBV-like outburst.
During the outburst, a total of $>0.01~{\rm M}_{\odot}$ of the 
outer layers of the star was ejected into space at a speed of 
$(1300 \pm 300)~{\rm km~s}^{-1}$, and heated to UV and X-ray emitting temperatures
as the outgoing shock hit this layer at a distance of $\sim 10^{16}$~cm
from the site of the explosion.

The extraordinary properties of \sn\ represent considerable theoretical challenges
regarding the modeling of Type Ib/c SN photospheres when additional (and perhaps dominating)
contributions from CSM interaction, especially cool dense shells between the 
ejecta and the pre-SN wind, have to be taken into account. 

\medskip
We gratefully acknowledge financial support of the NSF and 
NASA, especially the {\it Swift} and {\it Chandra} Guest Investigator Programs.

\end{document}